\documentclass[aps,prl,twocolumn]{revtex4}

\usepackage{times}
\usepackage{graphicx}
\usepackage{float}
\usepackage{latexsym,amsmath,amssymb,bm,euscript}
\usepackage{color}
\usepackage{subfigure}
\usepackage{epstopdf}
\usepackage[colorlinks=true,linkcolor=blue,citecolor=blue]{hyperref}
\usepackage{type1cm}
\usepackage{ulem}

\usepackage{appendix}

\begin{document}
\title{Interaction-Driven Spontaneous Quantum Hall Effect on Kagome Lattice}
\author{W. Zhu$^1$, S. S. Gong$^{2}$, T. S. Zeng$^{1}$, L. Fu$^{3}$ and D. N. Sheng$^1$}
\affiliation{$^1$Department of Physics and Astronomy, California State University, Northridge, California 91330, USA}
\affiliation{$^2$National High Magnetic Field Laboratory, Florida State University, Tallahassee, Florida 32310, USA}
\affiliation{$^3$Department of Physics, Massachusetts Institute of Technology, Cambridge, Massachusetts 02139, USA}

\begin{abstract}
Non-interacting topological states of matter can be realized in band insulators
with intrinsic spin-orbital couplings as a result of the nontrivial band topology.
In recent years, the possibility of realizing novel interaction-driven topological phase
has attracted a lot of research activities, which
may significantly extend the classes of topological states of matter.
Here, we report a new finding of  an interaction-driven spontaneous quantum Hall effect (QHE) (Chern insulator)
emerging in an extended fermion-Hubbard model on kagome lattice.
By means of the state-of-the-art density-matrix renormalization group,
we expose universal properties of the QHE
including time-reversal symmetry spontaneous breaking and quantized Hall conductance.
By accessing the ground state in large systems,
we demonstrate the robustness of the QHE against finite-size effects. 
Moreover, we map out a phase diagram and identify two competing charge density wave phases by varying interactions,
where transitions to the QHE phase are determined to be of the first order.
Our study provides a ``proof-of-the-principle''  demonstration of interaction-driven
QHE without requirement of external magnetic field or magnetic doping.
\end{abstract}




\maketitle


\textit{Introduction.---}
The integer quantum Hall effect (QHE) \cite{Klitzing1980}, the first example of the topological states of matter,
exhibits the topological property encapsulated by a quantized Hall conductance
in the presence of strong external magnetic fields \cite{Laughlin1981}.
The integer QHE is attributed to the topology of
electronic structures characterized by the first \textit{Chern number} \cite{Thouless1982,Avron2003}.
In particular, the connection between quantized Hall conductance and Chern number enables the QHE without applying any magnetic field,
or quantum anomalous Hall (QAH) effect.
Haldane pioneered an explicit lattice model realizing QAH effect \cite{Haldane1988},
by introducing a staggered magnetic flux to break time-reversal symmetry (TRS).
Remarkably, recent advances in experimental techniques have led to realization of QAH effect
in magnetically doped topological insulator via the interplay of ferromagnetism and spin-orbit coupling \cite{Xue2013,CXLiu2016} and cold atom systems via 
synthetic gauge fields \cite{Jotzu2014}.

The strong correlation between electrons is considered to be
another mechanism for inducing nontrivial topological states of matter,
which has attracted a lot of attention for their fundamental importance and potential applications.
One of the questions under debate is whether a QAH effect can be purely induced
by interactions in a microscopic model with
time-reversal symmetry and trivial  band topology.
In fact, there have been a series of proposals along this direction
\cite{Wen1989,Wen1991,CJWu2004,Raghu2008,KaiSun2009,YZhang2009,WenJun2010,Weeks2010,Kurita2011,Ruegg2011,YRan2011,Aoki2015,
Stefan2011,James2014,Tsai2015,Dora2014,Rahul2010,Pujari2016},
where the common wisdom is that strong interactions could generate circulating currents
and break time-reversal symmetry (TRS) spontaneously \cite{Wen1989,CJWu2004}.
The above picture is supported by different mean-field \cite{Raghu2008,WenJun2010,KaiSun2009,YZhang2009,Kurita2011,Weeks2010,Ruegg2011,YRan2011,Aoki2015}
or low energy renormalization group studies \cite{Stefan2011,James2014,Tsai2015,Dora2014,Rahul2010,Pujari2016}.
However,
unbiased numerical simulations such as exact diagonalization (ED) and density matrix renormalization group (DMRG)
studies found other competing states as the true ground states in previously proposed systems with Dirac points, such as honeycomb lattice models
\cite{Noel2013,Motruk2015,Capponi2015}.

The similar ideas were extended to other lattice systems with quadratic band crossing points (QBCP),
such as  checkerboard \cite{KaiSun2009}, kagome \cite{KaiSun2009,WenJun2010}, diamond \cite{YZhang2009} and Lieb lattice \cite{Tsai2015}.
On one hand, in the low-energy continnum theory and renormalization-group analysis,
QBCP is predicted to be unstable towards a QAH phase for infinitesimal repulsive interactions
 \cite{Stefan2011,James2014,Tsai2015,Dora2014,Rahul2010,Pujari2016} by opening an exponential small energy gap for excitations.
On the other hand, to our best knowledge,
previous attempts in microscopic models did not find evidences to support these mean-field results neither
\cite{Nishimoto2010,Pollmann2014}.
On the theoretical side, several existing challenges hinder the discovery of topological QAH phase in realistic lattice models.
Numerical studies indicate that
instead of triggering the desired TRS spontaneously breaking,
strong interactions tend to stabilize competing solid orders by breaking translational or rotational lattice symmetry.
Thus, the putative topological phase is usually preempted by various ordered states \cite{Noel2013,Motruk2015,Capponi2015}.
In addition, lacking of efficient method for detecting such kind of exotic phases makes
the problem  technically challenging.
For example, it is nontrivial to detect spontaneously TRS breaking,
as the TRS partners usually tend to couple together on small system sizes.
When the interaction is weak,
numerical simulations can hardly distinguish a possible semi-metal phase with gapless excitations
from an insulating topological state with a small excitation gap.
Therefore, the simple concept of realizing interaction-induced QAH phases
remains illusive for realistic electron systems.

\begin{figure}[t]
\includegraphics[width=0.45\linewidth]{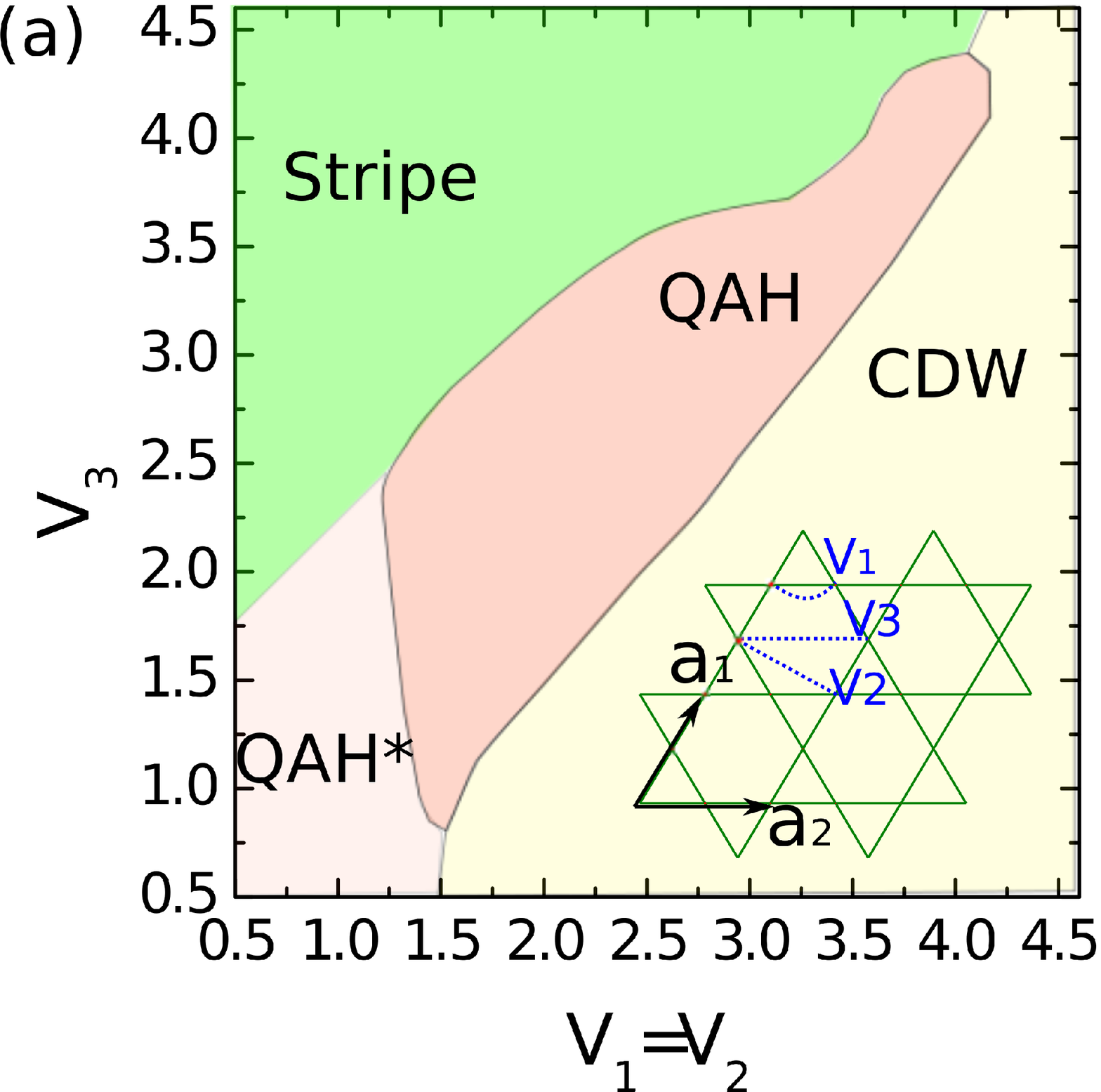}
\includegraphics[width=0.45\linewidth]{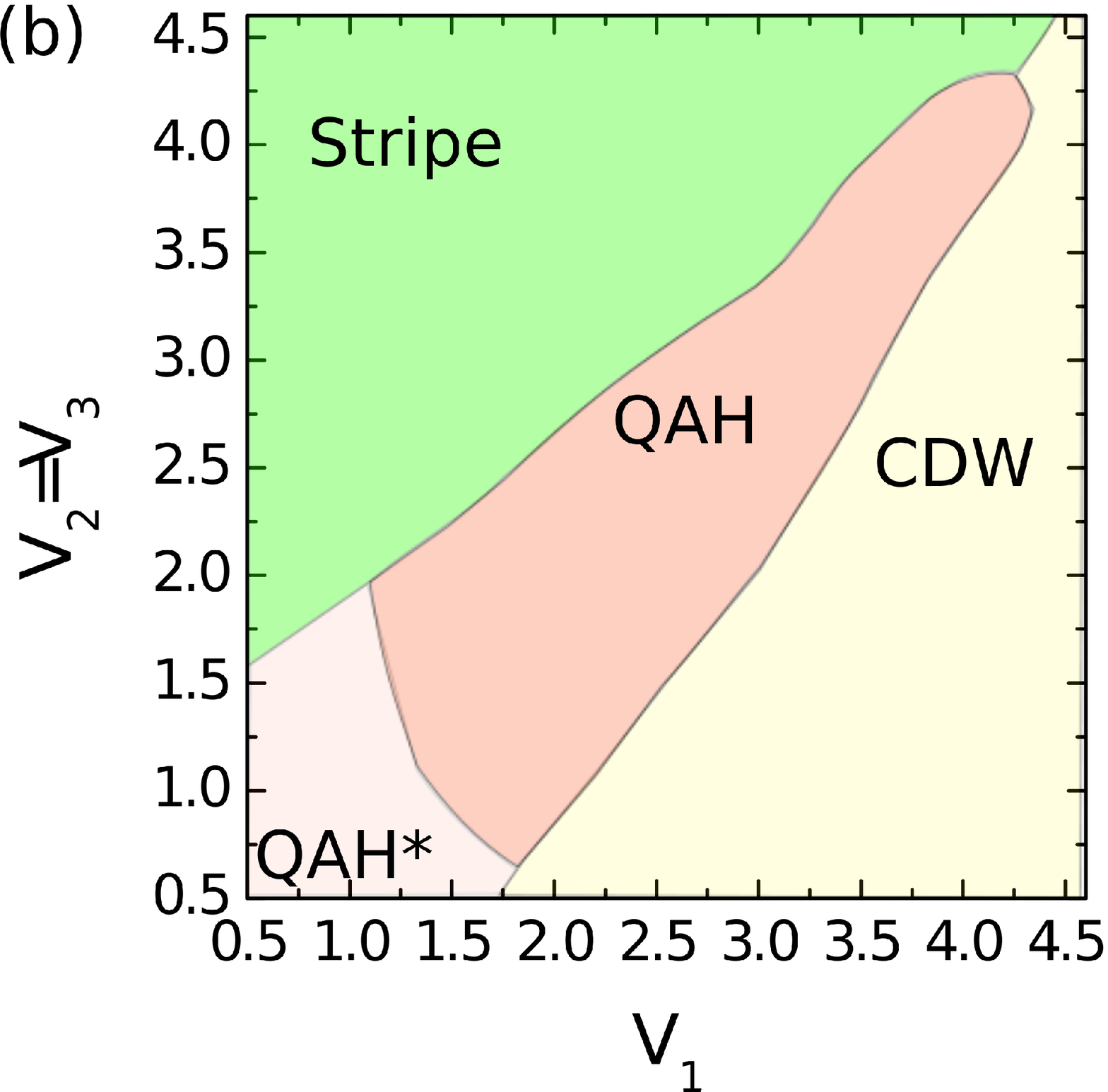}
\includegraphics[width=0.45\textwidth]{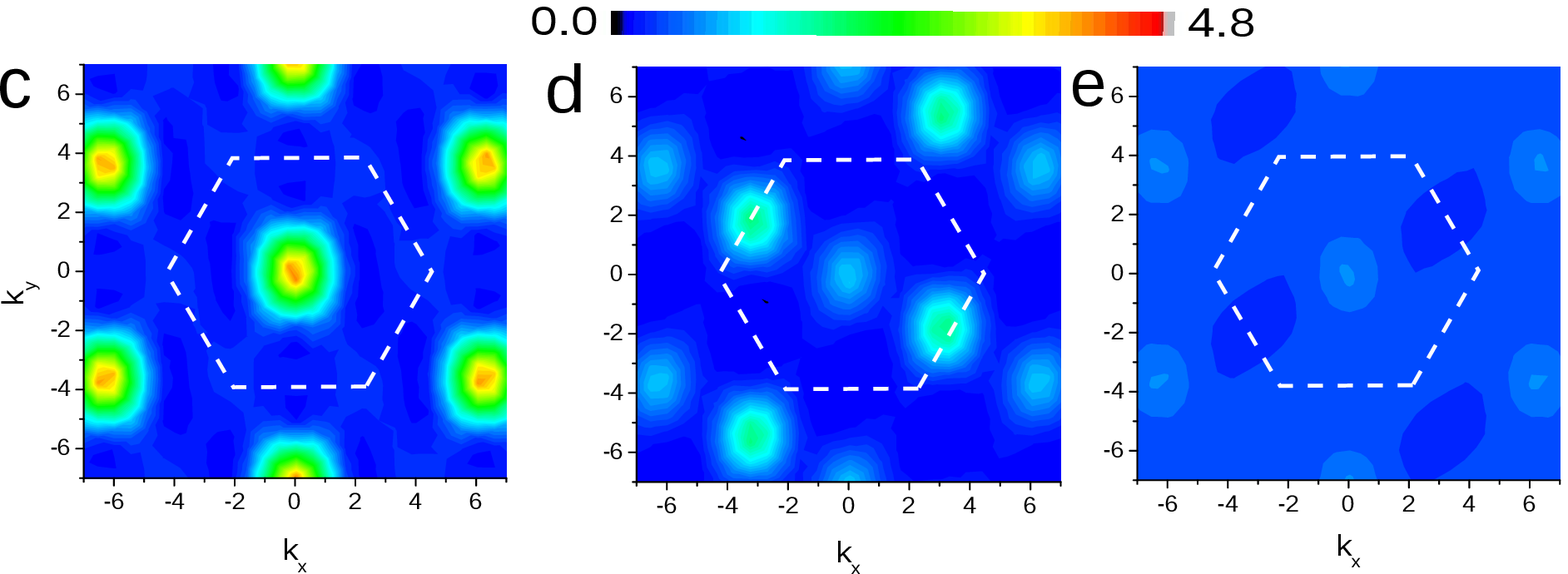}
\caption{Phase diagram of an extended fermion Hubbard model (Eq. \ref{eq:ham})
plotted in (a) $V_1=V_2$ and $V_3$ parameter space and
in (b) $V_1$ and $V_2=V_3$ parameter space,
obtained by DMRG calculations on cylinder of circumference $L_y=6$.
The QAH phase is characterized by the long-ranged current-current correlations
and integerly quantized Hall conductance.
The phase boundary between QAH phase and other phase is  determined by
the emergent loop current which signals TRS spontaneously breaking.
The contour plots of static density structure factor for:
(c) charge density wave $q=(0,0)$ phase, (d) stripe phase and (e) QAH phase.
The white dashed line shows the first Brillouin zone.
}  \label{phase}
\end{figure}

In this paper, we will address related  issues
and explore the possibility of interaction-driven QAH phase
by systematic  numerical simulations with applying  the state-of-the-art DMRG method for detecting topological states of matter
 \cite{SWhite1992,McCulloch2008,Cincio2013,HCJiang2012,YCHe2014,SSGong_SR,Zaletel_JSM,Grushin2015,WZhu2015}.
We will study  a kagome lattice model at one-third filling
and establish that the QAH phase can indeed be generated through  engineering electron interactions.
To be specific, as illustrated in the inset of Fig. \ref{phase}(a),
the model has nearest-neighbor hopping amplitude $t=1$ as energy scale,
as well as density-density repulsive interactions on first, second and third nearest neighbors,
described by the following Hamiltonian:
\begin{eqnarray}
H= &&t\sum_{\langle\mathbf{r}\mathbf{r}^{ \prime}\rangle} \left[c^{\dagger}_{\mathbf{r}^{\prime}}c_{\mathbf{r}}+\textrm{H.c.}\right]
+ V_1 \sum_{\langle\mathbf{r}\mathbf{r}^{ \prime}\rangle} n_{\mathbf{r}} n_{\mathbf{r}^{\prime}}  \nonumber\\
&+& V_2 \sum_{\langle\langle\mathbf{r}\mathbf{r}^{ \prime}\rangle\rangle} n_{\mathbf{r}} n_{\mathbf{r}^{\prime}} +
V_3 \sum_{\langle\langle\langle\mathbf{r}\mathbf{r}^{ \prime}\rangle\rangle\rangle} n_{\mathbf{r}} n_{\mathbf{r}^{\prime}},
\label{eq:ham}
\end{eqnarray}
where $c^{\dagger}_{\mathbf{r}}$ ($c_{\mathbf{r}}$) creates (annihilates) a spinless fermion at site $\mathbf{r}$.
We focus on the total filling number $\nu=N_e/N_s=1/3$
($N_e$ is the total electron number and $N_s$ is the  number of the lattice sites).
In the non-interacting limit, the lowest flat band quadratically touches the second band at the $\Gamma$ point ($K=(0,0)$)
\cite{supple}, thus the system is gapless at $\nu=1/3$ and topological trivial.
For the same model with only considering the $V_1$ interaction,
earlier DMRG simulations did not find any TRS breaking states \cite{Nishimoto2010}.
In the presence of strong interactions, our main findings are summarized in the phase diagram Fig. \ref{phase}(a-b).
In the intermediate parameter region (labeled by red),
we find a robust QAH phase emerging with the TRS spontaneously breaking.
The QAH phase is featured  by a twofold ground state degeneracy on torus geometry,
arising from two sets of QAH states with opposite chiralities.
The topological nature of the QAH states are characterized by the integer quantized Chern numbered $C=\pm 1$
for the TRS breaking states with the opposite chiralities, respectively.
In addition,  we also show that the QAH phase is neighboring with several solid phases which all respect TRS:
a stripe phase and a charge density wave phase,
both  demonstrating  distinctive Bragg peaks in their density-density  structure factors (Fig. \ref{phase}(c-d)).
On the contrary, the QAH phase displays a structureless feature (Fig. \ref{phase}(e)) in the structure factor,
indicating the absence of the space-group symmetry breaking.
Finally, reducing $V_1$, $V_2$ and $V_3$  simultaneously,
we find a parameter region shaded by light red in the left bottom corner in the phase diagram Fig. \ref{phase}(a-b),
which is likely a weaker QAH phase (labeled as QAH$^*$) as we discuss more details later.
Our results not only provide ``smoking gun'' evidences of interaction-driven topological phases,
but also shed insights into gapped and symmetry-broken phases in kagome lattice
\cite{Nishimoto2010,WZhu2015b}.

In order to study the ground state phase diagram in the $\{V_1,V_2,V_3\}$ parameter space,
we implement the DMRG algorithm \cite{SWhite1992,McCulloch2008}
combined with ED,
both of which have been proven to be powerful and complementary tools for studying realistic models
containing arbitrary strong and frustrated interactions \cite{Cincio2013,HCJiang2012,YCHe2014,SSGong_SR,Zaletel_JSM,Grushin2015,WZhu2015}.
We study large systems up to $L_y=6$ unit cells
and keep up to $M=4800$ states to guarantee a good convergence (the discarded truncation error is less than $2 \times 10^{-6}$).
We take advantage of the recent development in DMRG algorithm
by  adiabatically inserting flux to probe the TRS  spontaneous  breaking and
the topological quantized Hall conductance (see \cite{supple} for computational details)
\cite{SSGong_SR,YCHe2014,Grushin2015,WZhu2015}.

\begin{figure} 
\includegraphics[width=0.45\linewidth]{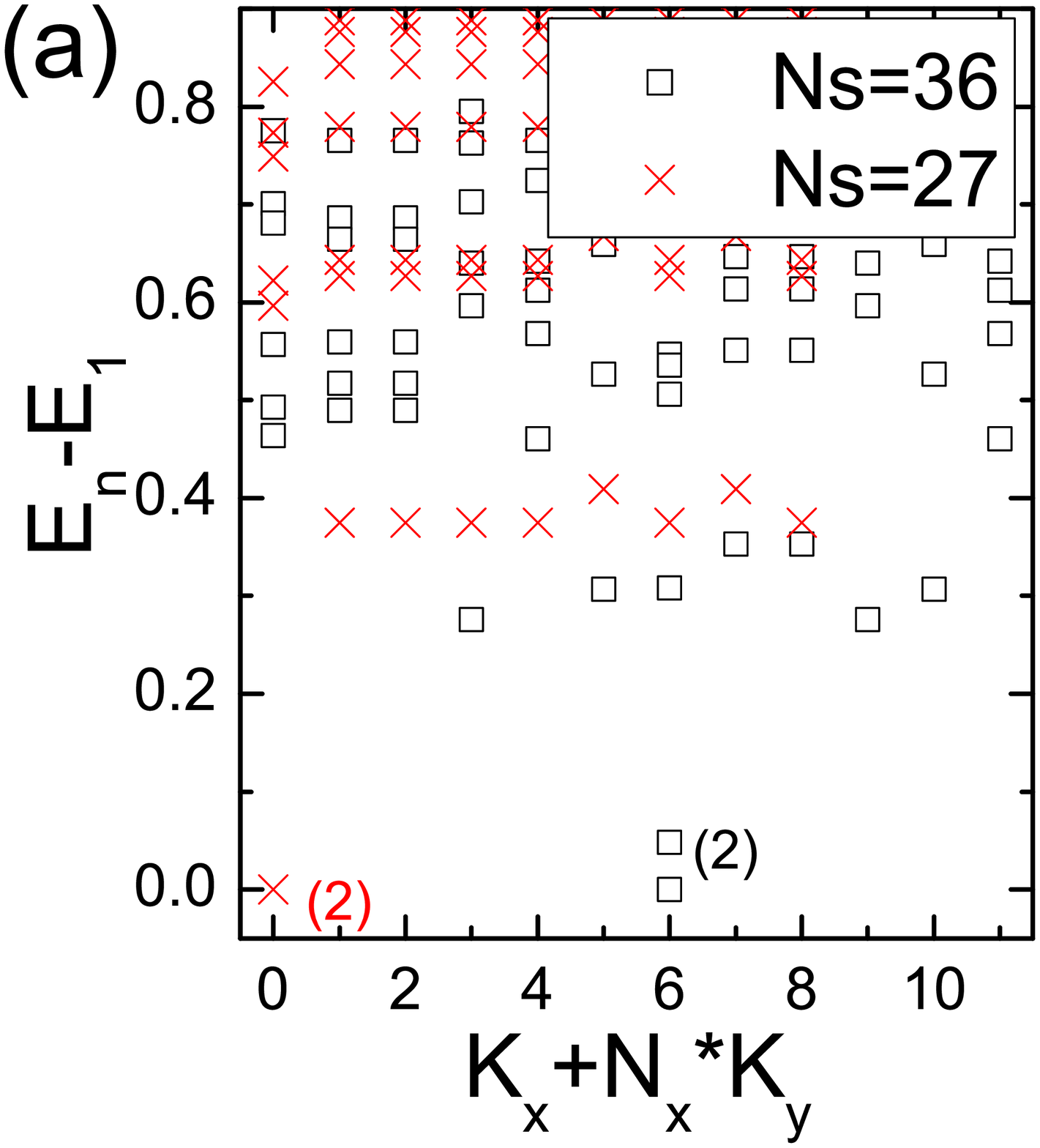}
 \includegraphics[width=0.45\linewidth]{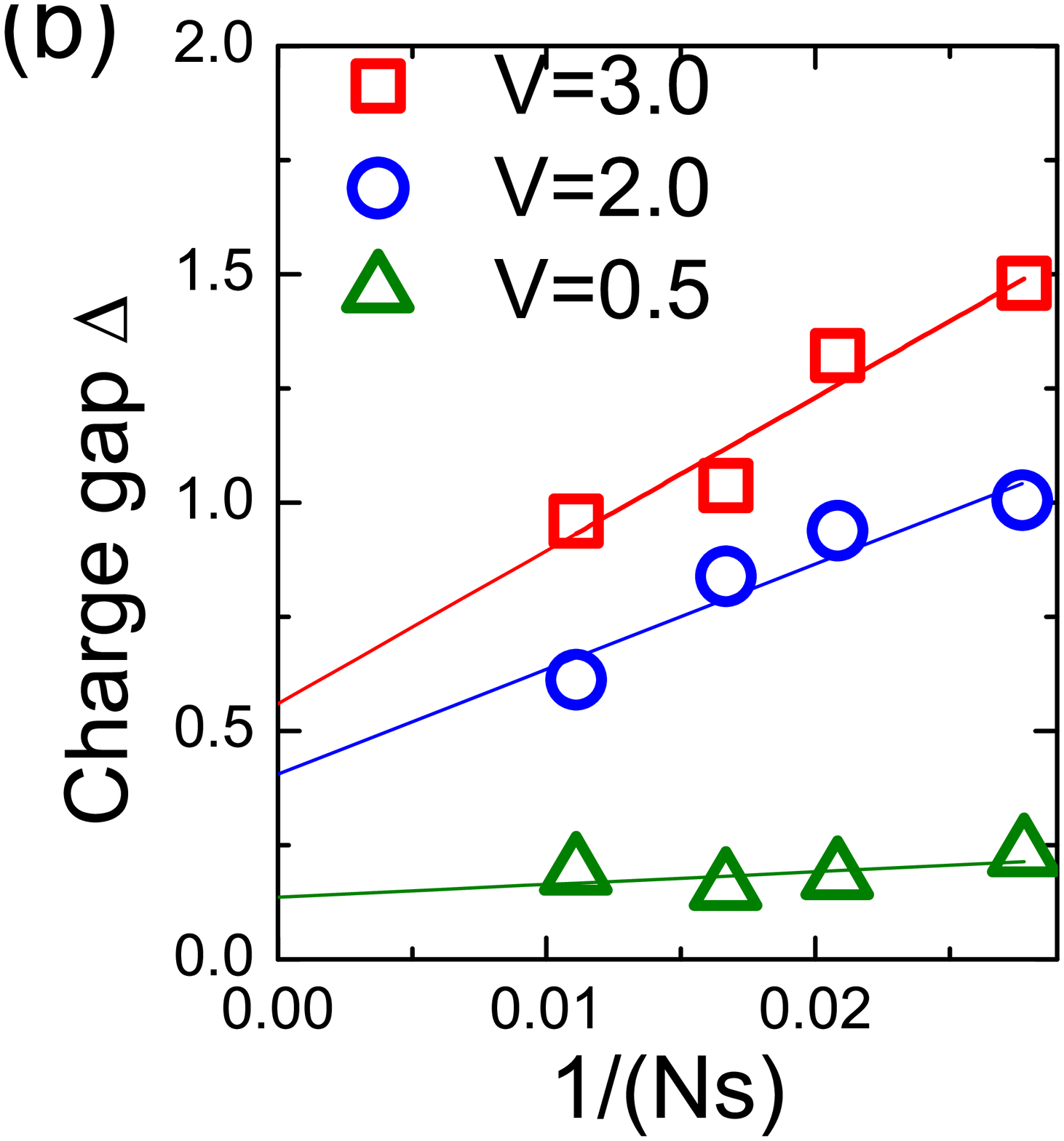}
\caption{(a) Energy spectra from ED versus momentum quantum numbers $(K_x,K_y)$ on the $N_s=3\times 3\times 3=27$ (red cross)
and $N_s=3\times 3\times 4=36$ (black square) sites cluster, by setting $V=V_1=V_2=V_3$ and $V=3.95$.
The ground state degeneracy are labeled by numbers.
(b) 
 Finite-size extrapolations of charge gap $\Delta(N_s)=E(N_s,N_e-1)+E(N_s,N_e+1)-E(N_s,N_e)$ obtained by DMRG
 ($E(N_s,N_e)$ the ground state energy on $N_s=3\times N_x \times N_y$ with $N_e$ electrons)
 on  several lattice clusters: $3\times 3\times 4$, $3\times 4\times 4$, $3\times 4\times 5$,
 $3\times 5\times 6$.
} \label{energy_gap}
\end{figure}

\textit{Energy Spectrum and Double Degeneracy.---}
The emergent QAH phase on a finite torus system is expected to host a two-fold ground state degeneracy,
representing  two TRS spontaneously breaking states with the opposite chiralities as TRS partners to each other.
To examine  this property for the  model systems, 
we first investigate the low-energy spectra based on ED calculation.
As shown in Fig. \ref{energy_gap} (a), we find  two near degenerating  ground states in energy spectra
for both $N_s=27$ and $36$ clusters \cite{note2}, which are separated from the excited levels by a finite energy gap.
Importantly, the ground states never mix with excited levels with varying the twisting boundary conditions,
signaling the robustness of excitation gap (see \cite{supple}). Moreover,
a stable topological phase is expected to be protected, not only by excitation gap,
but also by nonzero charge gap. In Fig. \ref{energy_gap}(b),
we also calculated the charge gap $\Delta(N_s)$
as a function of $1/N_s$, where the
finite-size scalings indicate a nonzero charge gap for QAH phase.

\begin{figure} [b]
\includegraphics[width=0.42\linewidth]{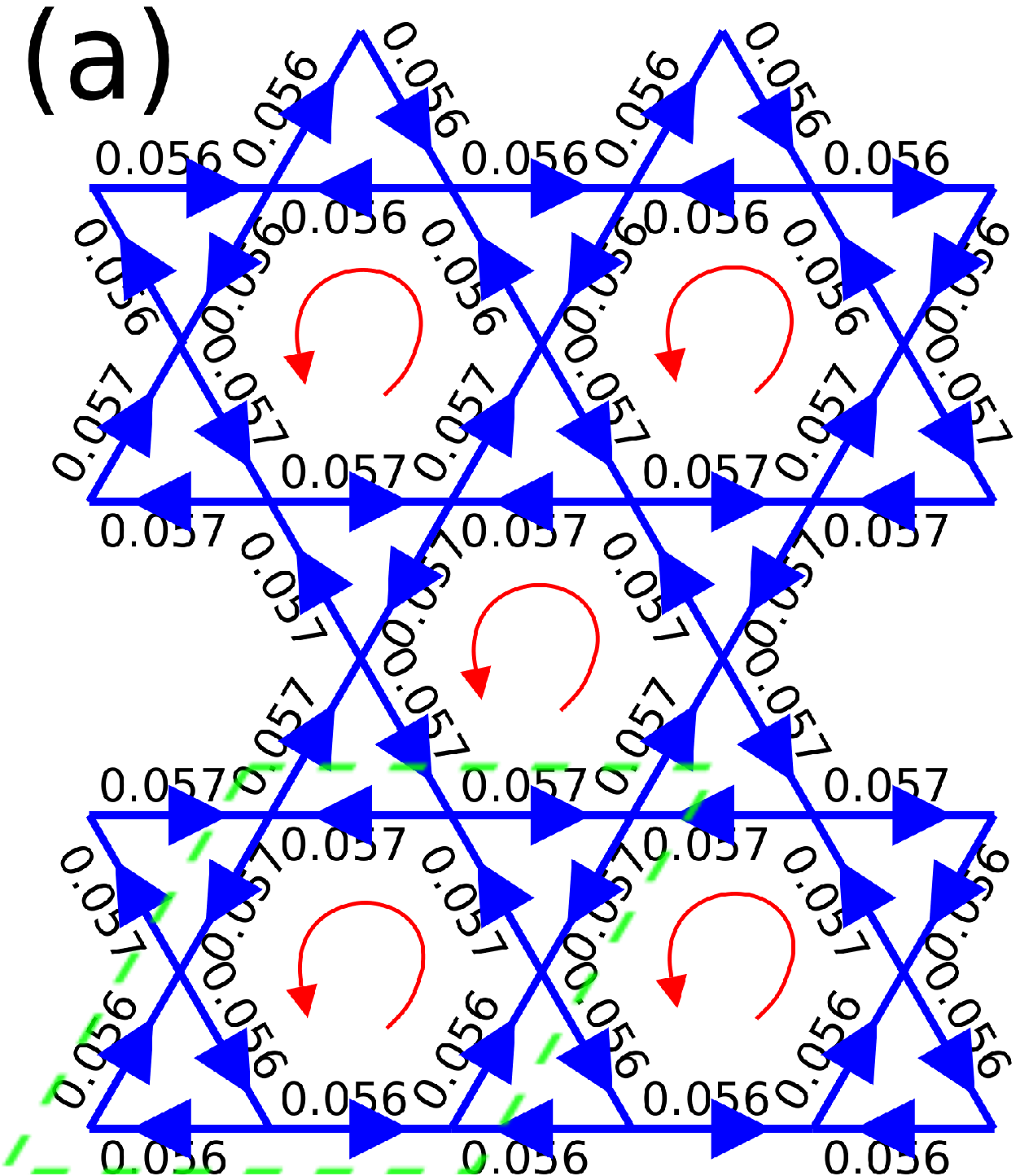}
\includegraphics[width=0.5\linewidth]{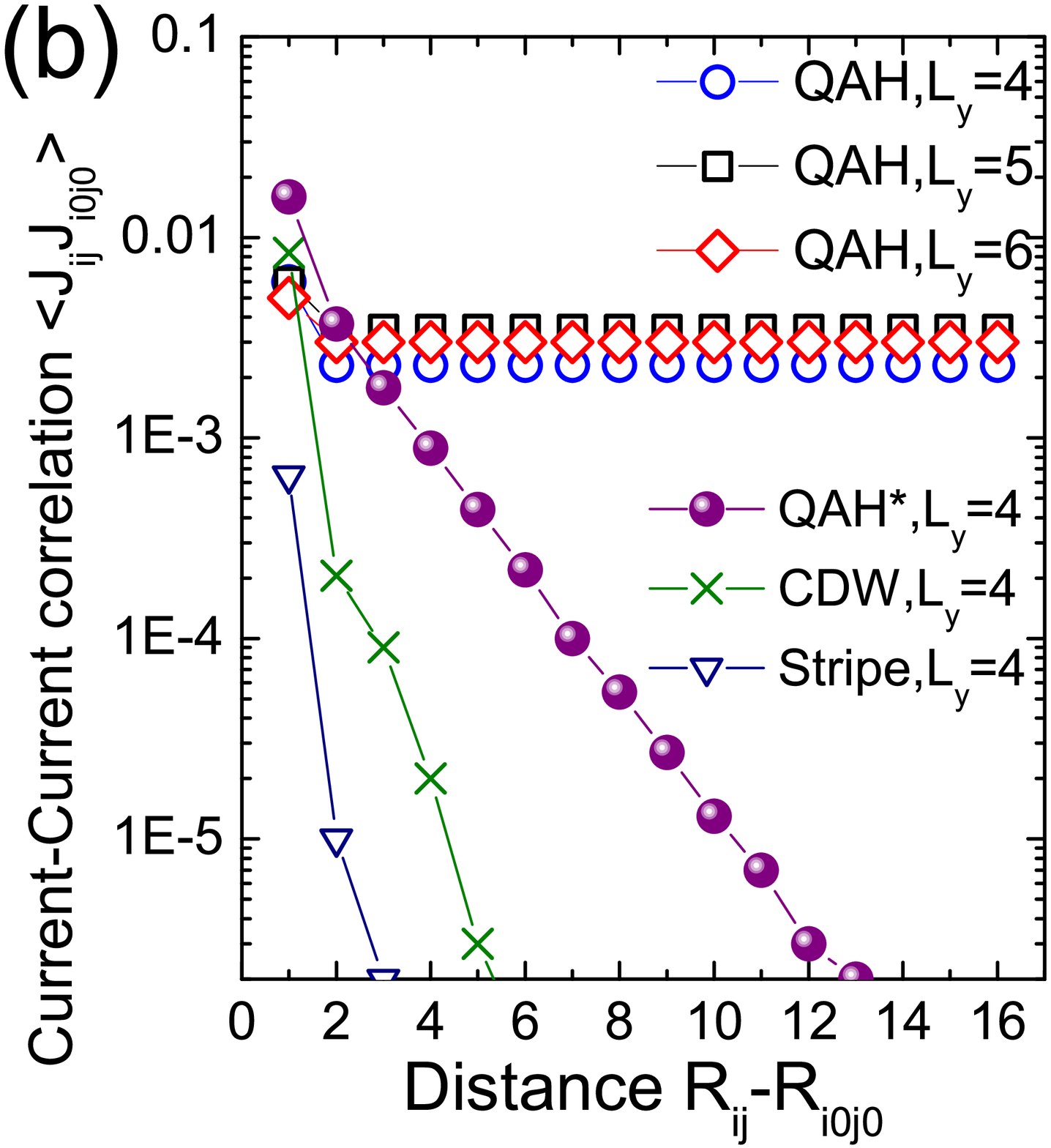}
\caption{(a) Real-space plot of emergent current pattern $\mathcal{J}_{ij}$
(we only show a section of three columns  on $L_y=4$ cylinder),
for  $|\Psi^{L} \rangle$ with left chirality.
The width of the bond is proportional to the absolute value (shown on the bond as a number) and
arrows point  to current directions. The red arrow indicates the current direction in each hexagon.  
(b) Log-linear plot of current-current correlations $\langle \mathcal{J}_{ij}\mathcal{J}_{i_0j_0}\rangle$
versus distance $R_{ij}-R_{i_0j_0}$ 
for QAH phase with system width $L_y=4$ (blue square), $L_y=5$ (black square) and $L_y=6$ (red square).
All correlations demonstrate long-range order (and they are also positive) for the QAH phase,
while $\langle \mathcal{J}_{ij}\mathcal{J}_{i_0j_0}\rangle$ decay exponentially in stripe phase
and charge density wave phase.
The real-space plot of current-current correlations is shown in Ref. \cite{supple}. } \label{current}
\end{figure}

\textit{Time Reversal Symmetry Spontaneously Breaking  and Emergent Loop Current.---}
To investigate the possible TRS \textit{spontaneously} breaking of the ground states,
we turn  to larger systems on the cylinder geometry and obtain the
ground states  by implementing DMRG calculation.
Indeed, we obtain two TRS breaking states $|\Psi^{L(R)}\rangle$
by random initializations of wavefunctions in DMRG simulations \cite{Cincio2013},
which are  degenerating in energy as expected (as the TRS partner to each other).
Here we label different groundstates by their chiral nature,
where $L$ ($R$) stands for ``left-hand'' (``right-hand'') chirality.
The corresponding TRS spontaneously breaking of $|\Psi^{L(R)}\rangle$ can be obtained by measuring emergent currents
$\mathcal{J}_{ij}=i\langle \Psi^{L(R)}|c^{\dagger}_i c_j-c^{\dagger}_j c_i|\Psi^{L(R)}\rangle$
between two nearest-neighbor sites $(i,j)$.
As shown in Fig. \ref{current} (a),
local current pattern $\mathcal{J}_{ij}$ uniformly distributes
(arrow representing direction of current),
which excludes the possibility of bond modulated local orders.
Most importantly,
local currents pattern form loop structure circulating
in the anti-clockwise direction in each hexagon for $|\Psi^L\rangle$
(We have checked that the TRS partner $|\Psi^{R}\rangle$ hosts clockwise loop current).
Interestingly, 
the staggered magnetic flux in each unit cell (enclosing one hexagon and two triangular) averages out to zero,
exactly matching the expectation of constructed  model for QAH effect \cite{Haldane1988,KaiSun2009}.

Moreover, we also calculate the current-current correlation functions $\langle \mathcal{J}_{ij}\mathcal{J}_{i_0j_0}\rangle$
in Fig. \ref{current}(b) ($(ij)$ is the bond parallel with the reference bond $(i_0j_0)$ and the distance measured by $R_{ij}$).
We compare $\langle \mathcal{J}_{ij}\mathcal{J}_{i_0j_0}\rangle$ for QAH phase with different system widths.
We find long-range correlations for all system widths  $L_y=4,5,6$.
The current correlations keep stable with increasing $L_y$, indicating
the TRS spontaneously breaking is robust against finite-size effects.
In contrast, 
the current correlations decay exponentially for stripe phase and charge density wave phase,
revealing the TRS preserving in solid phases.
As last, we notice that QAH$^*$ phase  can develop a relatively weaker current correlation,
albeit it shows sharply decaying correlation in short-range distance.

\begin{figure} 
 \begin{minipage}{0.5\textwidth}
 \centering
 \includegraphics[width=0.45\linewidth]{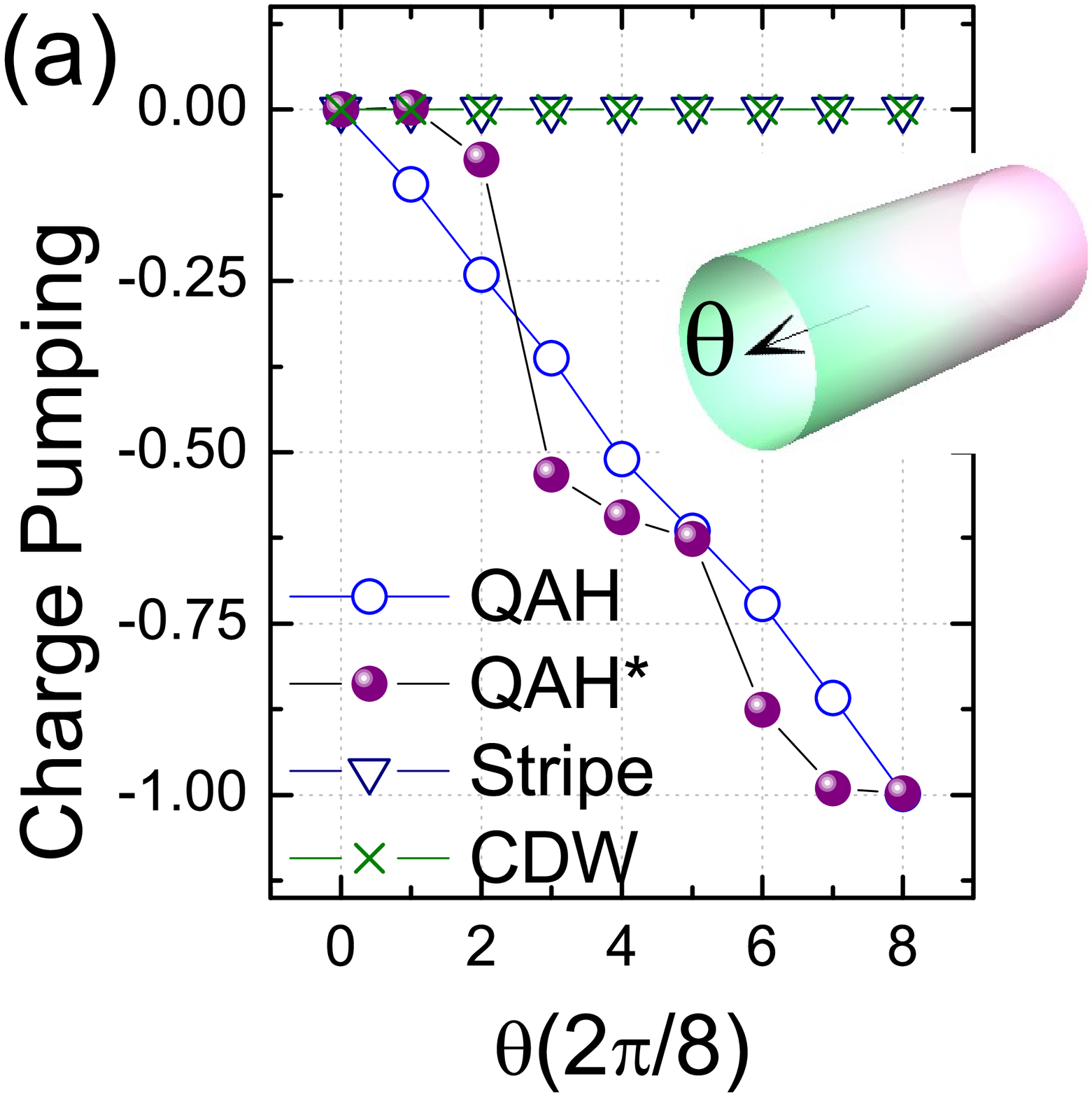}
 \includegraphics[width=0.45\linewidth]{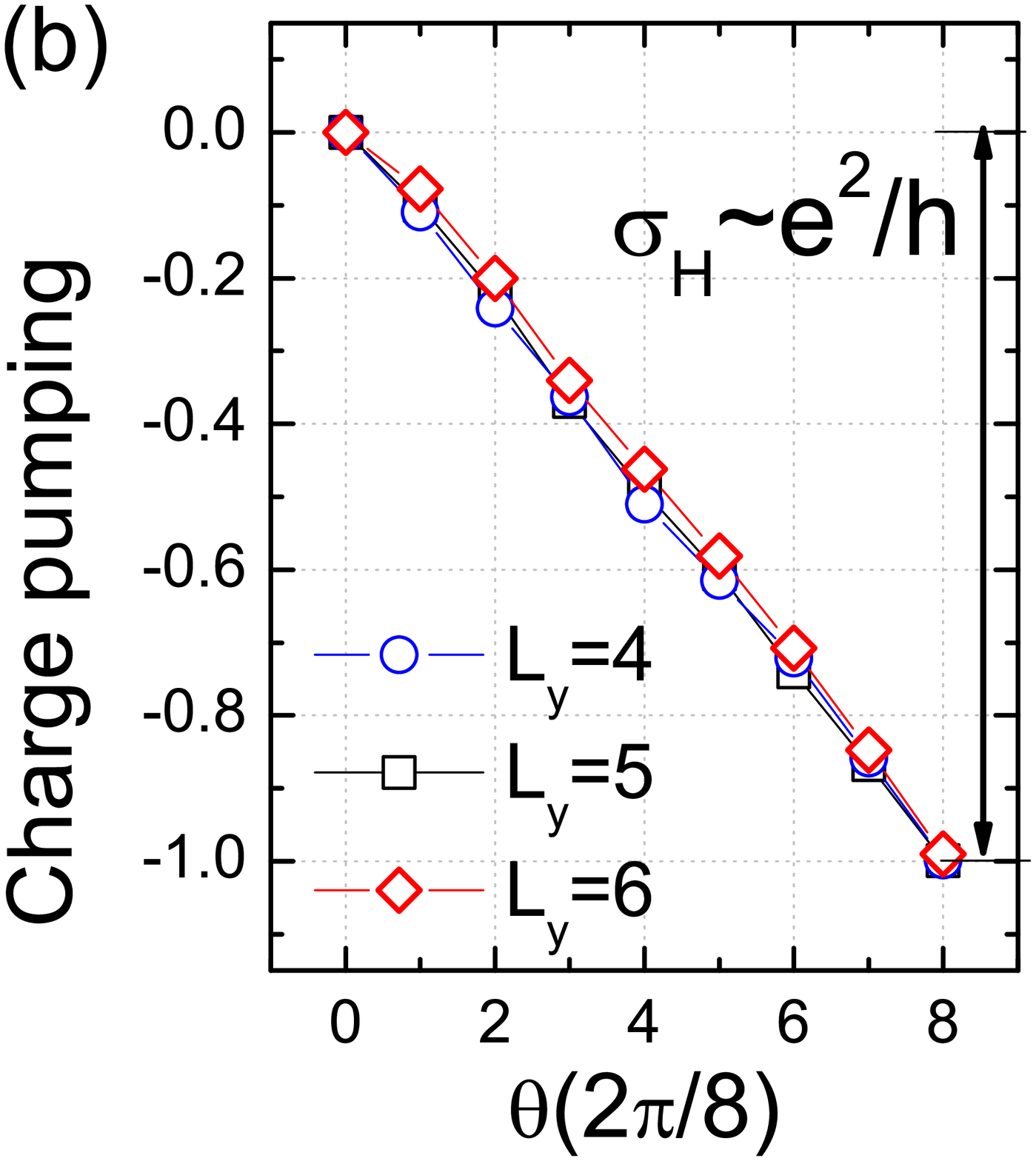}
 \end{minipage}
 \caption{Hall conductance $\sigma_H$ obtained by Laughlin flux insertion gedanken experiment,
 where $\sigma_H$ equals to the charge transfer $\Delta Q$ from one edge to the other edge.
(a) Net charge transfer $\Delta Q$ for QAH phase (blue circle), stripe phase (navy triangular),
charge density wave (green cross) and QAH$^*$ phase (purple dot). The system size is $L_y=4$ cylinder.
Inset cartoon illustrates adiabatically threading a $U(1)$ charge flux in the hole of cylinder.
(b) Net charge transfer $\Delta Q$ of QAH phase for different system sizes $L_y=4$ (blue circle),
$L_y=5$ (black square) and $L_y=6$ (red diamond).}
\label{chern}
\end{figure}

\textit{Quantized Hall conductance.---}
To uncover the topological nature of the QAH phase,
we perform a numerical flux insertion simulation on cylinder system
\cite{SSGong_SR,Zaletel_JSM}
to determine the quantized Hall conductance $\sigma_H$.
This simulation follows the idea of Laughlin gedanken experiment for interpreting integer QHE
\cite{Laughlin1981,Sheng2003},
where an integer quantized charge will be pumped from one edge to the other edge by inserting
a $U(1)$ charge flux $\theta$ in the hole of the cylinder.
At the DMRG side, we adiabatically increase the inserted flux $\theta$ and use the converged
wavefunction for smaller $\theta$ as the initial state for the increased $\theta$ to achieve
adiabatical evolution of the ground state \cite{SSGong_SR,supple}.
The  Hall conductance can be computed by 
$\sigma_H=\frac{e^2}{h}\Delta Q |^{\theta=2\pi}_{\theta=0}$
\cite{SSGong_SR,Zaletel_JSM}, where the net charge transfer $\Delta Q(\theta)$
can be calculated from the net change of the total charge in the half system:
$\Delta Q(\theta)=Tr[\hat \rho_L(\theta) \hat Q]$ ($\hat \rho_L$ the reduced density matrix of left half cylinder).
As expected, in Fig. \ref{chern}(a), the obtained $\sigma_H$ of QAH phase
takes nearly quantized value $\sigma_H\approx -1.00 e^2/h$ (for $|\Psi^{L}\rangle$)
by threading a flux quantum $\theta=0\rightarrow 2\pi$.
We also checked that the TRS partner $|\Psi^{R}\rangle$ hosts $\sigma_H\approx 1.00 \frac{e^2}{h}$.
In comparison, both the stripe phase and charge density wave do not respond to the inserted flux,
therefore have exactly zero Hall conductance $\sigma_H=0$ (Fig. \ref{chern}(a))
consistent with the  trivial topology  of  these states. 
Furthermore, we examine the stability of the topological quantization on finite-size systems.
In Fig. \ref{chern}(b),
we show the Hall conductance $\sigma_H$ of the QAH phase on cylinder system with widths $L_y=4,5,6$,
all of which give nearly quantized value $\sigma_H\approx -1.00 \frac{e^2}{h}$,
supporting that the  QAH phase is stable in the thermodynamic limit.

\textit{Phase transition.---}
We  address the nature of the quantum phase transitions between the QAH phase and
other phases (see \cite{supple}).
We utilize several quantities, such as groundstate wavefunction fidelity,
which are expected to signal the sensitivity of the wavefunction
with varying interacting parameters.
Moreover, we have also inspected several order parameters related to TRS
and translational symmetry spontaneously breakings\cite{supple}, respectively.
Based on these studies, we find that   transitions between the QAH phase and the stripe phase as well as charge density wave phase
are the first order ones, with evidences from step-like change in wavefunction overlap
and order parameters \cite{supple}.

Even though the QAH phase is shown to be remarkably robust in the phase diagram,
we are less certain about the QAH$^*$ phase (sitting at the left bottom corner of Fig. \ref{phase}).
The QAH$^*$ phase does not develop  solid orderings  by examining the structure factor.
We do not observe TRS breaking (or long-ranged current correlations) in QAH$^*$ phase.
Interestingly, if we perform flux insertion (inducing TRS breaking explicitly),
we  observe a nearly quantized Hall conductance (Fig. \ref{chern}(a)),
while  the evolution of the pumped charge versus flux is not as smooth as the QAH phase.
Hence, we believe that the ground state of QAH$^*$ is QAH state, however with strong finite size effect. 
For the system sizes we can access,  it is a  superposition state of the two  QAH  states  with opposite chiralities.
We conjecture that TRS in QAH$^*$ phase may be breaking when the system size becomes very large,
while more definite conclusion is beyond the current limit of computational capability.
In Fig. \ref{phase}, the phase boundary between QAH and QAH$^*$ phase (dashed line) is
determined by examining if the TRS is breaking or not on $L_y=6$ cylinder without inserted flux.
Going closer to the weak interaction limit, in DMRG and ED calculations, we find the ground state
remains to evolve adiabatically from the QAH$^*$ phase without additional quantum phase transition,
which is consistent with the theoretical expectation of infinitesimal interaction inducing
QAH effect \cite{KaiSun2009, Stefan2011,James2014,Tsai2015,Dora2014,Rahul2010}.

\textit{Conclusion and Outlook.---}
We have presented convincing evidences of an interaction-driven
spontaneous quantum anomalous Hall (QAH) phase in an extended Fermion-Hubbard model
on kagome lattice at one-third filling through engineering interactions.
Our complete characterization of the universal properties of the QAH phase
includes ground state degeneracy, time-reversal symmetry (TRS)
spontaneously breaking, and the quantized Hall conductance,
all of which provide an unambiguous diagnose of a QAH phase.
Such an exotic state had been sought after for a long time, however,
its existence in a microscopic model has remained elusive
until now.
Our {\color{red}} current results offer a ``proof-of-the-principle'' demonstration of
the spontaneous QAH purely driven by interactions, without the need of external magnetic field
or  other mechanism of explicit TRS breaking.
We believe our  work will stimulate future research along a number of directions.
For example,
introducing additional degrees of freedom in simple models usually results in  richer behaviors,
hence our current model with  including spin or orbital degrees will provide a promising  playground for
synthesising and engineering other  exotic states,
such as an emergent quantum spin Hall effect \cite{Kane2005} without spin-orbital coupling.
Moreover, we note that the lowest energy band on kagome lattice is exactly flat \cite{supple,Wu2007,Wu2008},
one could imagine a nearly flat band with non-zero Chern number after the gap opening by interactions.
This would be quite significant since it may provide a platform to realize the fractional QAH phase
when such flat band is partially filled \cite{Titus2011,Tang2011,KaiSun2011,DNSheng2011}.
At experimental side,  based on the recently experimental development  of artificial kagome systems \cite{Jo2012},
we anticipate activities to realize and detect  the QAH state in ultracold atomic systems \cite{Jotzu2014}.

\textit{Acknowledgements.---}
We thank K. Sun and Y. Zhang for stimulating discussions.
This research is supported by
the DOE office of Basic Energy Sciences
under grants No. DE-FG02-06ER46305 (W.Z., D.N.S),
S.S.G is supported by the National High Magnetic Field Laboratory
(NSF DMR-1157490) and the State of Florida.
L.F. is supported by the
DOE office  of Basic Energy Sciences, Division of Materials Sciences and Engineering under award de-sc0010526.
We also acknowledge partial support from NSF grant DMR-1532249 for computational resource.

\textit{Note added.} During the final stages of the completion of this
manuscript, we became aware of a work claiming a QAH phase on a checkerboard lattice
based on ED calculation on small sizes \cite{ZYMeng2016}
(see also \cite{note1}).


\bibliographystyle{apsrev}
\bibliography{qahe_kagome}{}

\begin{thebibliography}{55}
\bibitem{note2}
For $N_s=3\times 3\times 3=27$ cluster,  the lowest states are exactly degenerating protected by the rotational symmetry.
Nevertheless, due to the finite-size effect, on system $N_s=36$ there is a finite splitting between lowest two states.
\end{thebibliography}




\clearpage
\vspace{40pt}
\begin{appendices}

\begin{widetext}
\tableofcontents

\section{A. Method} \label{method}
In this paper, the calculations are based on the density-matrix renormalization group (DMRG) algorithm on cylinder geometry \cite{SWhite1992,McCulloch2008}
and the exact diagonalization (ED) on torus geometry, both of which have been proven to be effective
and complementary  tools for studying realistic models with arbitrary strong and frustrated interactions.
The combination of ED and DMRG is powerful and complementary.
On one hand, in ED it is straightforward in identifying the ground state degeneracy.  
But with the exponential growing
of the Hilbert space, the accessible systems are limited to smaller system sizes,
for example, up to $N_s=36$ in this study. On the other hand, DMRG calcualtion allows us to obtain
accurate groundstates on much larger system sizes beyond the ED limit.
Moreover, DMRG calculation has great advantages
of probing ground states with spontaneous symmetry breaking and topological ordering
 \cite{Cincio2013,HCJiang2012,YCHe2014,SSGong_SR,Grushin2015,WZhu2015}.


\subsection{1. Details of DMRG Calculation}
We study the cylinder system with open boundaries in the x direction
and periodic boundary condition in the y direction.
The available system sizes are cylinders of circumference $L_y=4,5,6$ (in unit of unit cell).
For the largest system width ($L_y = 6$, with $12$ lattice constants in length), we keep up to $M=4800$ $U(1)$
states and reach the DMRG truncation error around $2 \times 10^{-6}$.
In addition, we also confirm that both infinite DMRG with system width $L_y=4,5,6$ unit cells
and finite DMRG for very long systems (up to $N_s=3\times 4 \times 36=576$ sites)
always obtain the same phase and the same topological quantization for QAH phase.


\subsection{2. Adiabatic DMRG and Quantized Chern Number}
We have used the numerical flux insertion experiment based on the adiabatical
DMRG simulation to detect the topological Chern number
of the bulk system \cite{SSGong_SR,Zaletel_JSM,WZhu2015}.
To simulate the flux $\theta$ threading in the hole of a cylinder, we impose the twist boundary conditions along
the y direction with replacing terms $c^{\dagger}_{\mathbf{r}^{\prime}}c_{\mathbf{r}}+h.c.\rightarrow
e^{i\theta_{\mathbf{r}^{\prime}\mathbf{r}}}c^{\dagger}_{\mathbf{r}^{\prime}}c_{\mathbf{r}}+h.c.$
for all neighboring $(\mathbf{r}, \mathbf{r}^{\prime})$ bonds
with hoppings crossing the y-boundary in the Hamiltonian (Eq. \ref{eq:ham}).
The charge pumping from one edge to the other edge can be computed from
$\langle \Delta Q(\theta)\rangle=Tr[\hat{\rho}_L(\theta) \hat{Q}(\theta)]$ \cite{Zaletel_JSM}, where
$\hat{Q}(\theta)$ is the $U(1)$ quantum number and $\hat{\rho}_L(\theta)$ is reduced density matrix of left half system.
Due to the quantized Hall response, the Chern number of ground state is equal to
the charge pumping by threading a $\theta=2\pi$ flux \cite{Laughlin1981}.
To realize the adiabatic flux insertion, we use the step of flux insertion as $\Delta \theta=0.25\pi$.


\subsection{3. Finite-Size Effect Analysis}   
In our calculations, we carefully checked the results are stable in the largest system size.
To be specific, in DMRG the largest available system size is $L_y=6$ cylinder for kagome lattice,
which is equivalent to $12$ lattice spacing.
We have double checked that, up to $L_y=6$ cylinder, Hall conductance of QAH phase
is quantized to be $\sigma_H \approx \pm 1.0 e^2/h$, as shown in the main text (Fig. \ref{chern}).
We also confirmed that, the emergent loop current is uniform and stable on $L_y=6$ system (Fig. \ref{fig:current_corr}).
These evidences support that the observed QAH phase is expected to be stable in the thermodynamic limit.

Moreover, when studying the topological order on cylinder geometry with finite width $L_y$,
the correlation length $\xi$ of the ground state offers a natural consistency check for the assumption
that the value of $L_y$ is large enough to be representative of the thermodynamic limit \cite{Cincio2013}.
The correlation length is defined by $\xi=-1/\ln\lambda_1$, $\lambda_1$ is the second largest eigenvalue of transfer-matrix \cite{McCulloch2008}.
If $L_y$ is much larger than $\xi$, we indeed expect finite-size effects to be very small.
Indeed, the condition $L_y>\xi$ is satisfied for QAH phase in our phase diagram, so that
our DMRG calculation offers a reliable and relevant results for thermodynamic limit.
Based on this measurements, we expect the finite-size effect should be small in our calculations.

\begin{figure}[b]
 \includegraphics[width=0.4\textwidth]{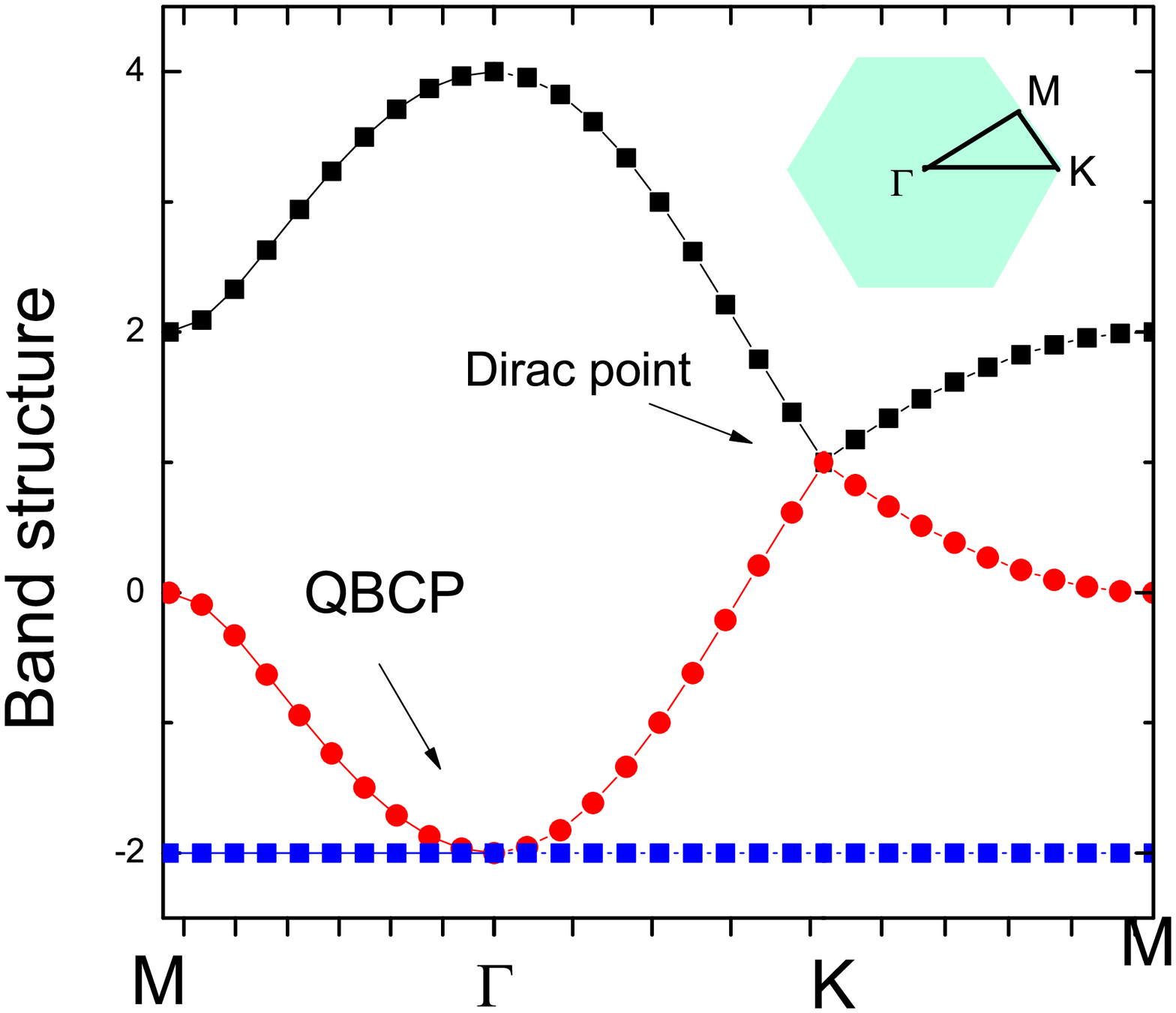}
 \includegraphics[width=0.5\textwidth]{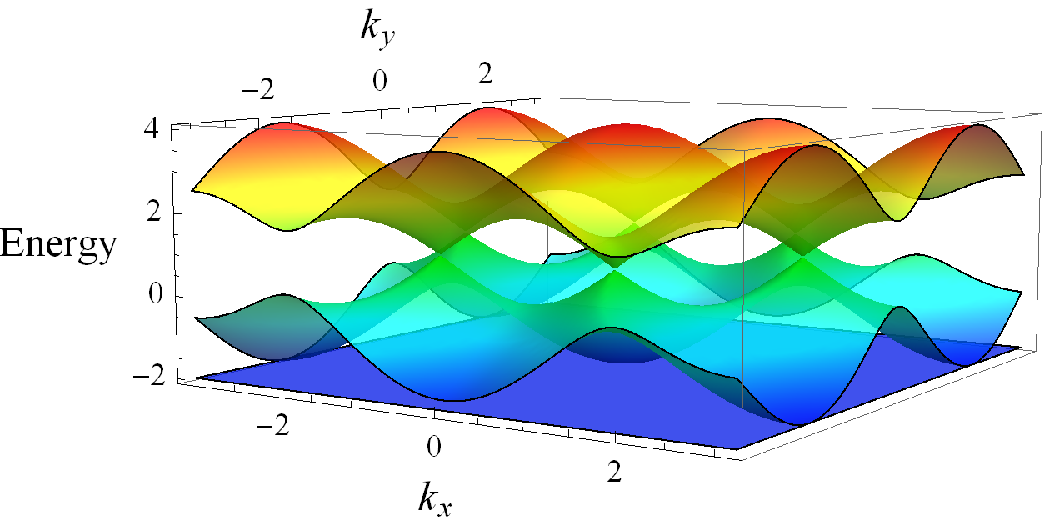}
 \caption{(left) Band structure of kagome lattice along symmetry point $\Gamma \rightarrow K \rightarrow M \rightarrow \Gamma$.
(right) Band structure of kagome lattice on whole Brillouin zone.
}\label{fig:band}
\end{figure}

\section{B. Tight-Binding Band Structure on Kagome Lattice}
%
We briefly discuss the tight-binding model of kagome lattice in the non-interacting limit:
$H_0=t\sum_{\langle ij \rangle} c^\dagger_{i} c_{j}$.
A section of the kagome lattice is shown in Fig.~\ref{phase} in the main text.
Kagome lattice shares an underlying triangular lattice and we choose the unit cell vectors to be
\begin{equation}
{\boldsymbol  a}_{1}=a(1,0)\quad {\rm and}\quad {\boldsymbol  a}_{2}=a(\frac{1}{2},\frac{\sqrt{3}}{2}),
\end{equation}
where $a=1$ is lattice constant. The kagome lattice has three sites in the unit cell.
The reciprocal lattice vectors are given by
\begin{equation}
{\boldsymbol b}_{1}=\frac{2\pi}{a} (1,\frac{-1}{\sqrt{3}})\quad {\rm and} \quad
{\boldsymbol b}_{2}=\frac{2\pi}{a} (0,\frac{2}{\sqrt{3}}).
\end{equation}
The first Brillouin zone forms a hexagon in momentum space for both lattices,
as shown in inset Fig. \ref{fig:band}.
%
%
The noninteracting energy dispersion for a nearest-neighbor tight-binding model can be obtained analytically.
In momentum space Hamiltonian becomes $H_0=\sum_{k}\Psi^\dagger_{k}{\cal H}^0_k \Psi_k$ with
$\Psi_{k}=(c_{1 k},c_{2 k}, c_{3 k})^T$ (The index $l=1,2,3$ in $c_{l,k}$ labels the three sublattices in a unit cell), and
\begin{equation}\label{hk0}
{\cal H}^0_k=2t\left(
\begin{array}{ccc}
  0 & \cos{k_1} & \cos{k_2} \\
  \cos{k_1} & 0 & \cos{k_3} \\
\cos{k_2} & \cos{k_3} & 0
\end{array}\right).
\end{equation}

By diagonalizing the Eq. \ref{hk0}, we get the dispersion relation:
\begin{equation}
\epsilon_{1}(\boldsymbol{k})=t+t A_{\boldsymbol{k}},\quad\epsilon_{2}(\boldsymbol{k})=t-t A_{\boldsymbol{k}},\quad \epsilon_{3}(\boldsymbol{k})=-2t.
\label{eq:kagome_disp}
\end{equation}
In Eq.~\eqref{eq:kagome_disp} we have defined
\begin{equation}
A_{\boldsymbol{k}}=\sqrt{3+2\cos k_{1}+2\cos k_{2}+2\cos (k_{1}-k_{2})},
\label{eq:Ak}
\end{equation}
where $k_1={{\boldsymbol k} \cdot \boldsymbol a}_1 $ and $k_2={{\boldsymbol k} \cdot \boldsymbol a}_2$.
There are two dispersing bands ($n=1$ and 2) and a flat band ($n=3$).
At filling fraction $\nu=2/3$, the two dispersing bands touch at two inequivalent Dirac points ($K_{\pm}$)
located at corners of the Brillouin zone
$\boldsymbol{K}_{\pm}=\pm(\boldsymbol{b}_{1}-\boldsymbol{b}_{2})/3$.
At filling fraction $\nu=1/3$,
the second band touches the flat band at the $\boldsymbol{\Gamma}$ point [${\boldsymbol K}=(0,0)]$.
This is a quadratic band crossing point (QBCP).
Including a bilinear interaction (etc. intrinsic spin-orbit coupling or dimer coupling)
can lead to formation of a gap opening at the Dirac points ($\nu=2/3$) or the QBCP ($\nu=1/3$).
In this paper, we focus on $\nu=1/3$ and explore the possibility of dynamically generating a topological phase from interactions
and study its competition with other broken-symmetry phases.

Besides the existence of a QBCP point and two Dirac points,
another interesting feature in band structure is the flat band (labeled by blue in Fig. \ref{fig:band}(left)).
There is indeed structural reason to guarantee this property on kagome lattice \cite{Wu2007,Wu2008}.
We can construct the local wavefunction $|\Psi_{i_h}>$ which is the state of a fermion on the $i_h$-th hexagon with an
effective momentum $k_r=\pi$ along the six sites of the hexagon.   One can easily show that each of these states is an eigenstate of noninteracting $H$ with
an eigenvalue $-2t$ as the net transfer from one hexagon to another is zero due to the destructive interference.
   There are $N_s/3$ of these linear independent states, which form the lowest energy flatband
by making translational invariant momentum eigenstates from them.

\begin{figure} 
 \includegraphics[width=0.7\linewidth]{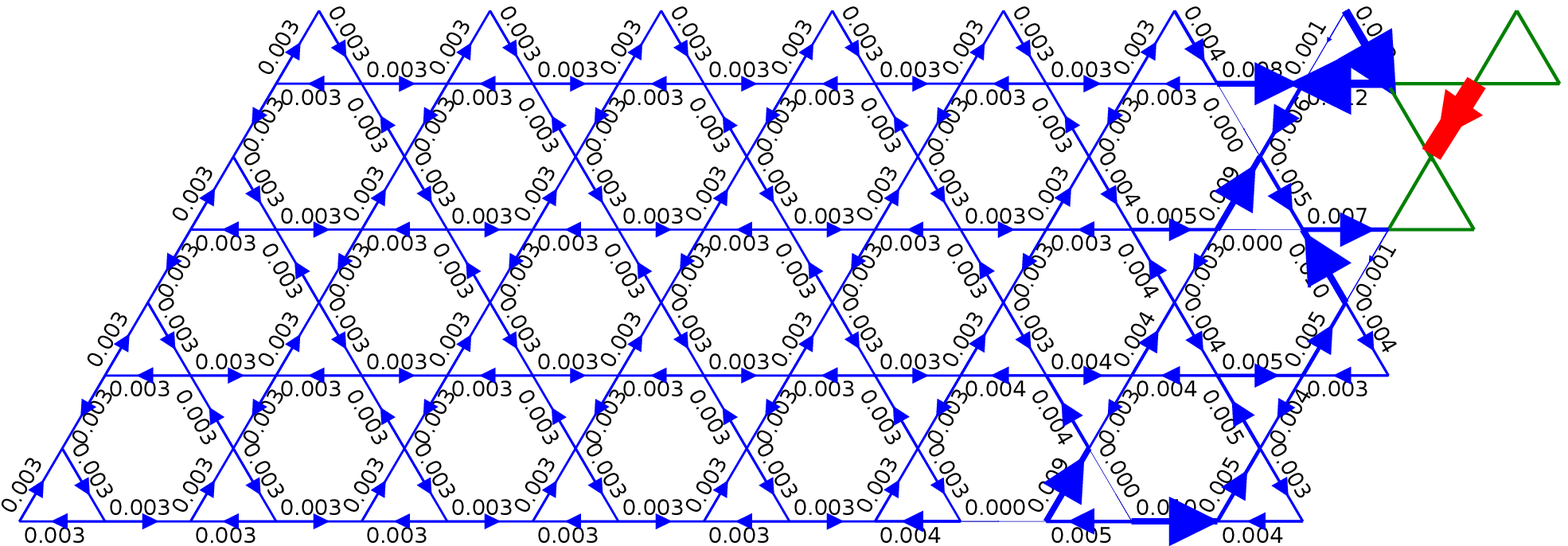}
 \includegraphics[width=0.9\linewidth]{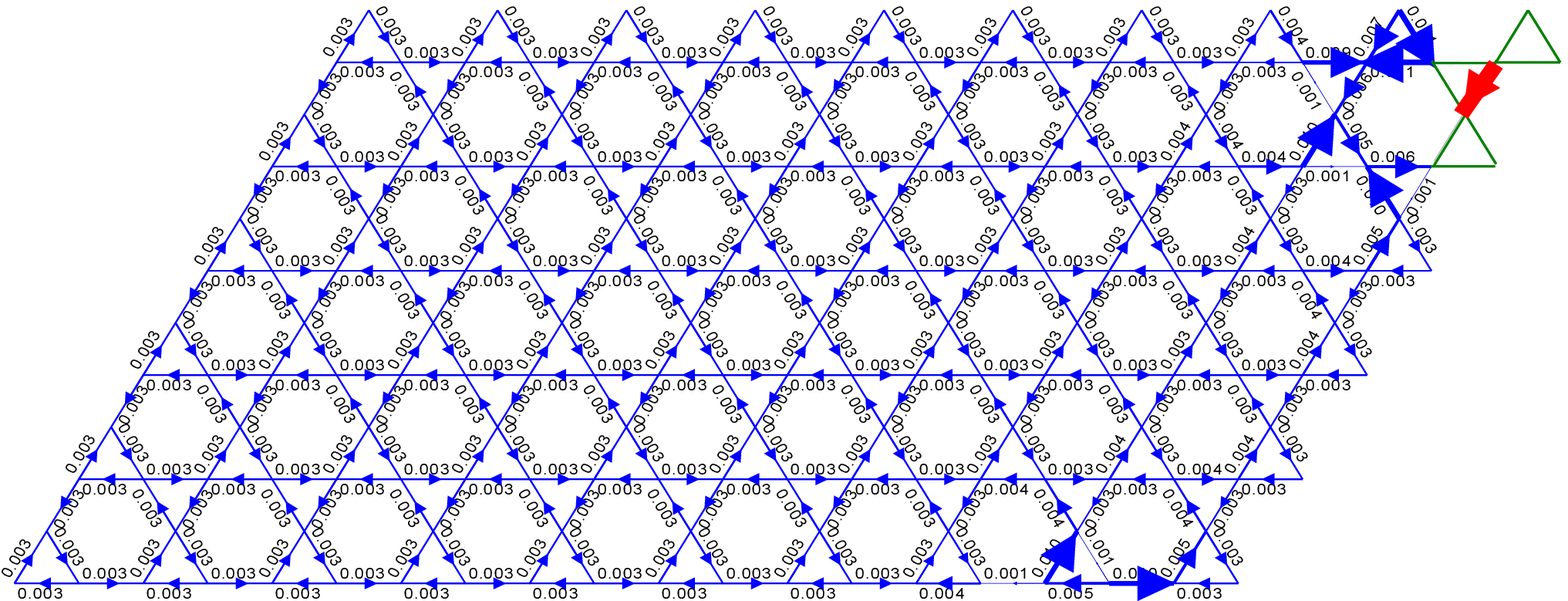}
 \caption{Real space plot of current-current correlations on (top) $L_y=4$ cylinder and (bottom) $L_y=6$ cylinder.
Here we show a sector of several columns near the reference bond (marked as red).
Width of bond is proportional to the absolute value (shown on the bond as a number) and
arrows correspond to current directions. The interaction parameters are $V_1=V_2=V_3=3.0$.
} \label{fig:current_corr}
\end{figure}

\section{D. Current-Current correlation}
When the system undergoes a transition to QAH phase, the corresponding TRS spontaneously breaking
can be obtained by drawing circulating currents. In order to check this picture,
we have demonstrate the local current pattern in the main text. In this section, we present further
numerical evidences from current-current correlations.
In Fig. \ref{fig:current_corr}, we plot the real-space distribution of current-current correlations between a reference bond (red)
and other bonds (The reference bond is put on the center of the cylinder and we show bonds on the left side of reference bond).
Intriguingly, all current correlations away from the reference bond
(distance larger than two lattice constant) match
the expectation of QAH phase, with the correct current direction on every bond.
And the current correlation distributes uniformly and reaches a value that can be
comparable to the square of order value (as shown in the main text).
Let us emphasize again, the results perfectly match QAH picture and no frustration is found.
To remove the finite-size effects, we also study the current-current correlation on $L_y=4,5,6$ system.
Up to the largest system ($L_y=6$), the current-current correlation shows long-ranged ordering, which indicates that
the current ordering is robust against finite-size effect. Thus, we expect the TRS spontaneously breaking is
a stable property in the thermodynamic limit.

\begin{figure} 
 \includegraphics[width=0.4\linewidth]{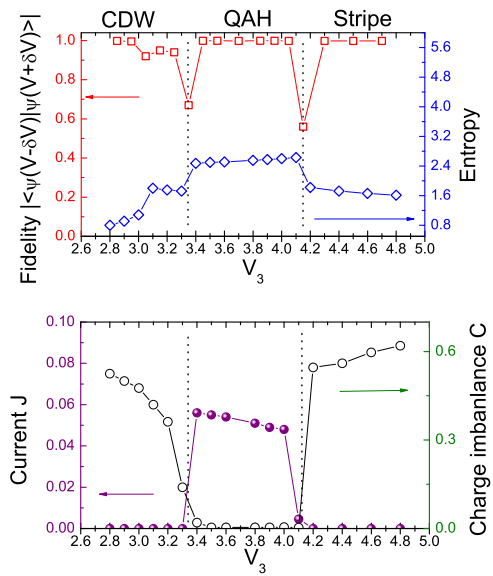}
 \caption{Topological transition between QAH phase and solid phases. We select one line in phase diagram
 by varying $V_3$ and setting $V_1=V_2=3.5$.
 Physics quantities includes: (a) Ground state wavefunction fidelity
 $|<\psi(V-\delta V)|\psi(V+\delta V)>|$ ($\delta V=0.1$)
 (red squares)  and entanglement entropy (blue squares),
 (b) current order parameter $J$ (purple dots) and charge imbalance order parameter $\Delta \rho$ (green circles).
 The calculations are performed on $L_y=4$ cylinder. } \label{fig:transition}
\end{figure}

\section{E. Quantum phase transitions} \label{QPT}
In order to uncover the nature of corresponding phase transitions between
QAH phase and solid phases, we
inspect several quantities that are expected to be sensitive to a phase transition,
such as order parameters related to TRS spontaneously breaking
$J=\frac{1}{N_s}\sum_{\langle ij\rangle} \varepsilon_{ij} \mathcal{J}_{ij}$ and
sublattice charge imbalance $\Delta \rho=\frac{1}{N_s}\sum_{i,\alpha} |\rho_{i,\alpha}-1/3|$,
where $\mathcal{J}_{ij}$ is current order between two nearest neighbor bond $\langle ij\rangle$
($\varepsilon_{ij}=\pm$ correspond to the expected QAH orientation) and
$\rho_{i,\alpha}$ is charge distribution on sublattice $\alpha$ in unit cell $i$.
Both quantities are expected to show a finite jump when crossing a first order transition.
We also calculate the groundstate wavefunction fidelity $F=|\langle\psi(V)|\psi(V+\delta V)\rangle|$
($V$ is some parameter in Hamiltonian),
which can faithfully describe the first-order transition or energy level crossing.

We show the results along a reference line,
by fixing $V_1=V_2=3.5$ and varying $V_3$.
In Fig. \ref{fig:transition}(a),
it is found that the wavefunction fidelity shows two dips around
$V_3\approx 3.2$ and around $V_3\approx 4.2$, indicating two transition points.
When looking at the entanglement entropy,
we also observe a quick jump around $V_3\approx 3.2$ and around $V_3\approx 4.2$.
Both of these two measurements signals a direct first order phase transition between
QAH phase and solid phases.
More information can be obtained by the order parameters $J$ and $\Delta \rho$.
The solid phase spontaneously breaks translational symmetry but preserves TRS ($\Delta \rho \neq 0, J=0$),
while QAH phase spontaneously breaks TRS but holds translational symmetry ($\Delta \rho=0$ and $J\neq 0$).
Indeed, in Fig. \ref{fig:transition}(b), it is observed that a nonzero and uniform loop current $J\neq 0$
associated with strong suppression of $\Delta \rho$ in the QAH region.
Based on these measurements, we determine the nature of transition between the QAH phase and
two solid phases to be first-order.

\begin{figure} 
 \includegraphics[width=0.45\linewidth]{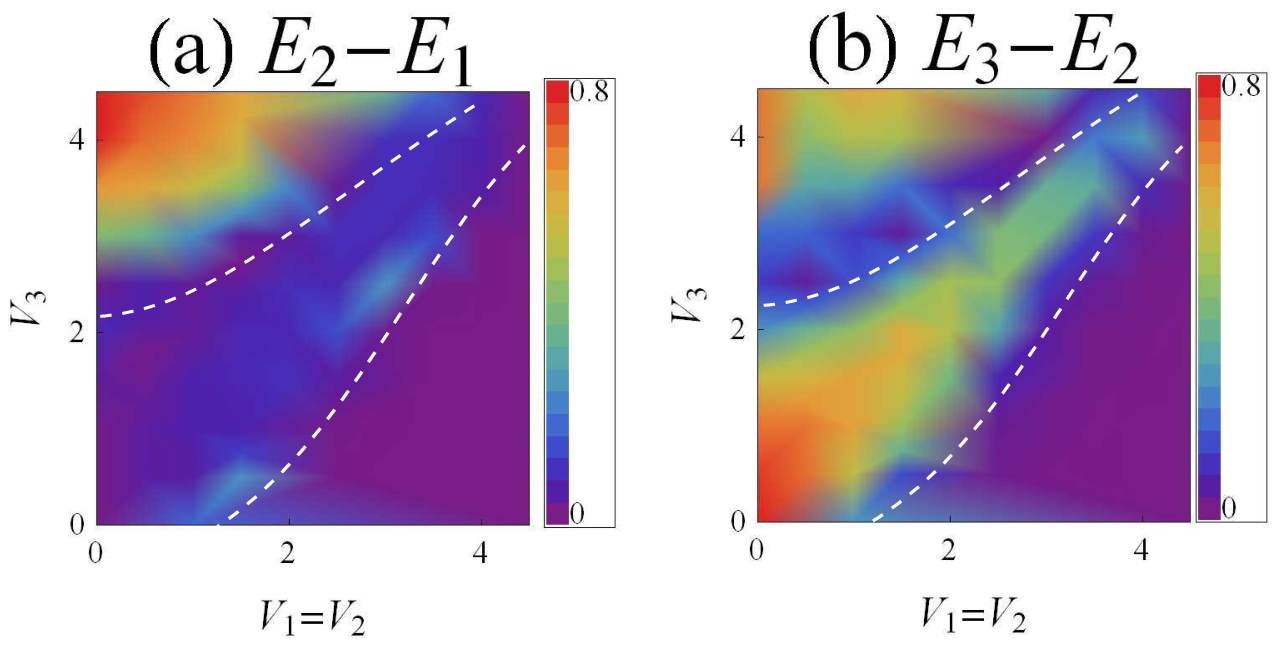}
 \includegraphics[width=0.3\linewidth]{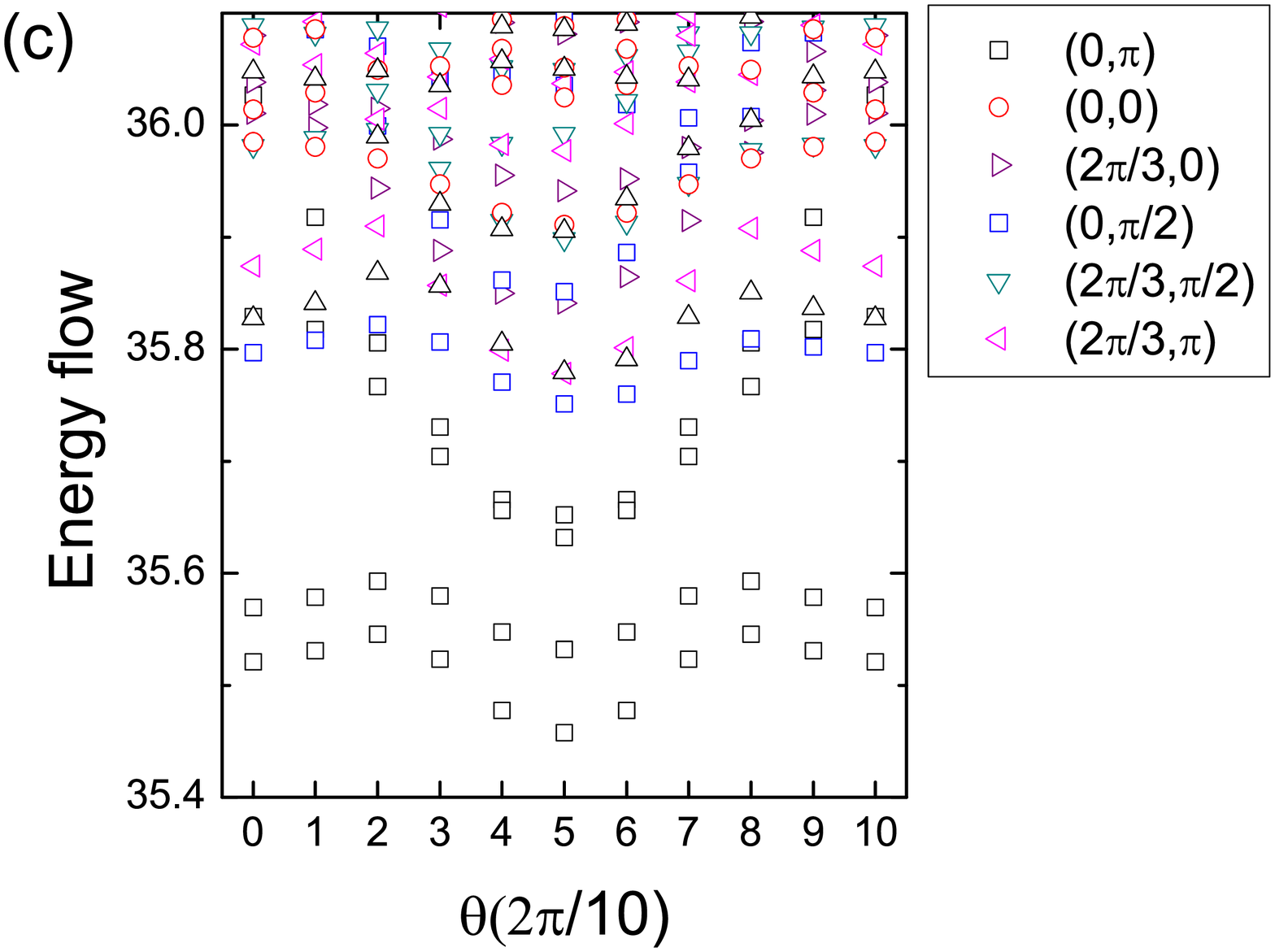}
 \caption{(a-b) Contour plot of energy difference (a) $E_2-E_1$ and (b) $E_3-E_2$,
 on $N_s=3\times 3\times 4$ cluster using ED. Here, $E_1,E_2,E_3$ stands for the lowest energy, second lowest energy
 and third lowest energy in momentum sector $(0,\pi)$.
 The white dashed line marks the phase boundary between QAH phase and CDW phase as well as stripe phase.
 (c) Energy spectra versus twisted boundary condition $\theta$, by setting $V_1=V_2=V_3=3.95$.
 Different momentum sectors are labeled by different symbols.
 } \label{fig:EDgap36}
\end{figure}

\section{H. Evidences from ED Calculation}
Here, we would like to present some more numerical evidence from ED calculation on torus system.
On torus geometry (periodic boundary condition), it is straightforward to study the
potential ground state degeneracy. Since QAH phase for a finite system
would appear as a twofold-quasidegenerate ground state, one for each chirality,
we expect to demonstrate this important property using ED.

On $N_s=3\times 3\times 4$ cluster, QAH phase is expected to have
twofold ground state degeneracy in momentum sector $K=(0,\pi)$,
while CDW phase have threefold ground state degeneracy in $K=(0,\pi)$.
In contrast, stripe phase hosts twofold degeneracy, one in $K=(0,0)$ and the other one in $K=(0,\pi)$.
Thus, we can distinguish the three phases by plotting energy gap $E_3-E_2$ and $E_2-E_1$
in momentum sector $K=(0,\pi)$.
As shown in Fig. \ref{fig:EDgap36}, the two phase boundaries are shown by
dips in $E_3-E_2$ (marked by white dashed line), indicating the energy level crossings.
To inspect the energy gap of QAH phase, we also introduce the twisted boundary condition and calculate the energy flow:
$\langle \vec r + N_x \hat x + N_y \hat y |\Psi_{\theta_x,\theta_y}\rangle=
e^{i(\theta_x+\theta_y)N_e}\langle \vec r|\Psi_{\theta_x,\theta_y}\rangle$,
($|\Psi_{\theta_x,\theta_y} \rangle$ the many-body state with boundary phase $\theta_x,\theta_y$),
which helps us distinguish robustness of two-fold ground state degeneracy.
Importantly, the two-fold ground states never mix with excited levels with varying the twisting boundary conditions (Fig. \ref{fig:EDgap36}(c)),
signaling the robustness of energy gap.

At last, we would like to point out, it is not trivial to characterize time-reversal
symmetry breaking in ED, because the ground states obtained from ED always preserve TRS symmetry due to periodic boundary condition.
As a result, a naive calculation of Chern number using ED will always give zero
since we would have a superposition of two ground states with different chiralities
that are related by reverting the sign of emergent staggered magnetic flux.
Therefore, a DMRG calculation is necessary and insightful in this problem,
where ground state with spontaneous symmetry breaking is favored on long cylinder due to minimal entropy rule
\cite{Cincio2013,HCJiang2012,WZhu2015b}.

\section{I. Mean-Field Analysis} \label{sec:mean_field}
In the main text, we present a quantum phase diagram obtained from DMRG calculations,
which is a full quantum mechanism method dealing with arbitrary strong and frustrated interactions
without any approximation on the quantum fluctuations. To gain a better understanding of our discovery,
it is helpful to study the mean-field phase diagram, as discussed below.


\begin{figure}[!htb]
 \includegraphics[width=0.3\linewidth]{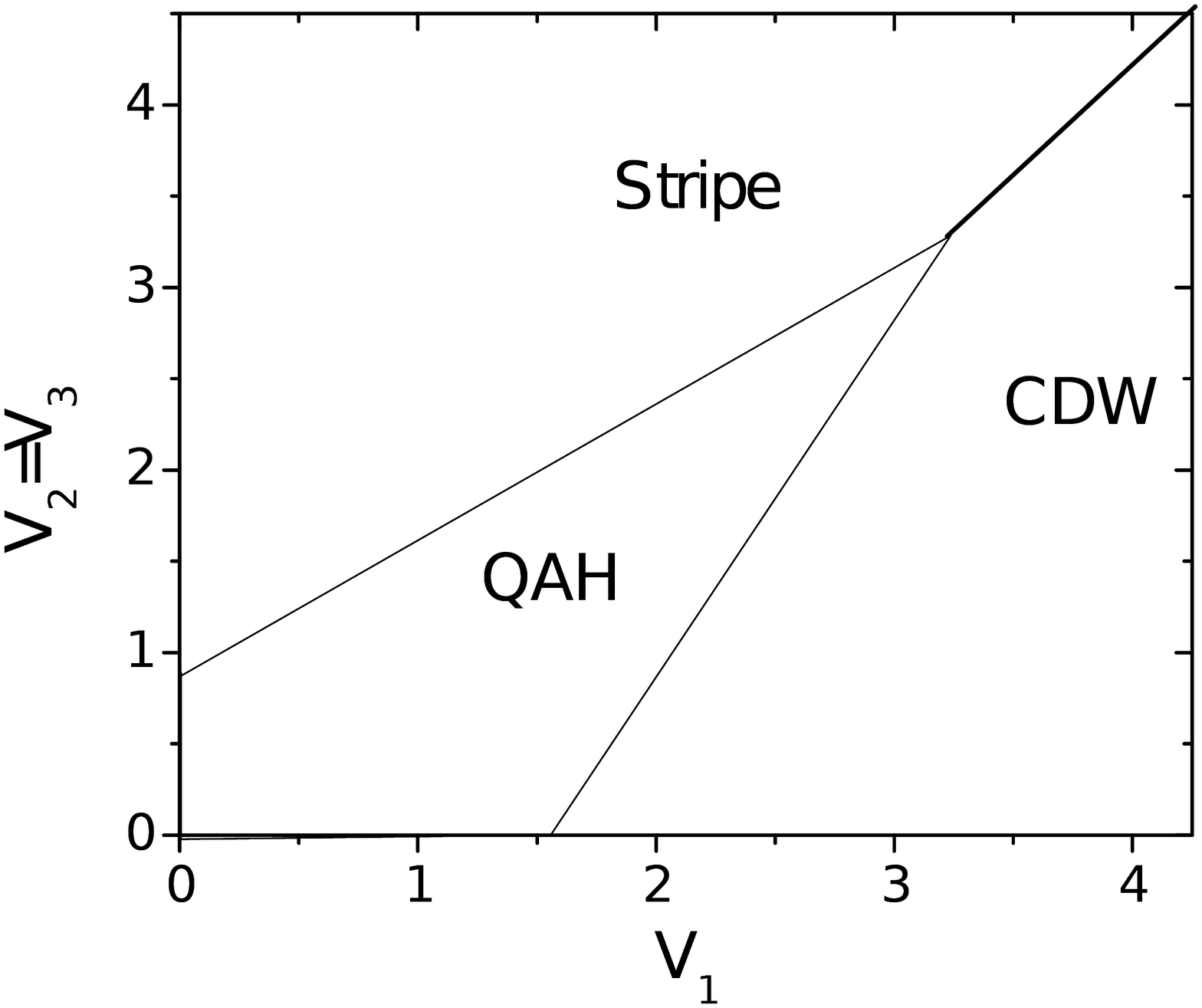}
 \includegraphics[width=0.5\linewidth]{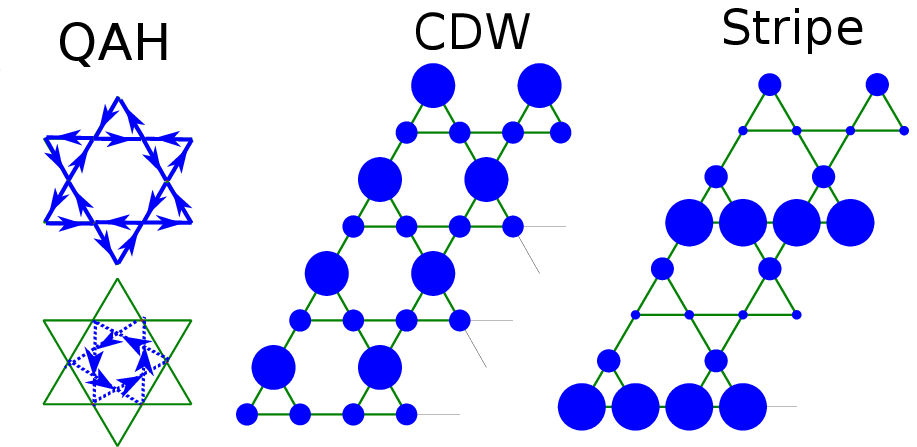}
 \caption{
 (a)Mean field diagram plotted in $V_1$ and $V_2=V_3$ parameter space. It is found a QAH phase
 sandwiched between stripe and charge density wave phase when interaction is not strong.
 (b)Cartoon picture of QAH phase, CDW phase and stripe phase.
 For QAH phase, there is flux pattern developed by $V_1$ and $V_2$ interactions in mean-field results.
 Charge distribution is uniform on all sites. The bond direction marks complex hopping between
 neighborhood sites. For CDW phase, the charge distribution on sublattices is non-uniform.
 For stripe phase, the translational symmetry is spontaneously breaking.
} \label{fig:Diagram}
\end{figure}

We present the phase diagram of the model obtained by mean-field approach.
As to the details of the mean-field decouplings and self-consistent equations,
we first replace the four fermion interaction terms (in Eq. (1) in main text)
with bilinear terms (Hartee-Fock approximation),
which can be interpreted as the additional hopping and potential energy terms in real space.
In the mean-field ansatz, we use a six-sites unit cell (enlarge original unit cell by a factor two),
and allow finite flux on nearest-neighbor (NN) hopping and also on next nearest-neighbor (NNN) hoppings.
Finally, we solve the mean-field equations self-consistently and take the solution that minimizes the free energy,
which is called unrestricted Hartee-Fock method in some literatures.

As shown the mean-field diagram in Fig. \ref{fig:Diagram}, besides
several topological trivial phases including a stripe phase, a charge density wave
and a semi-metal phase (non-interacting point), it is found an emergent QAH phase.
Mean-field analysis supports that interactions tend to favor
the QAH phase when interactions are relatively weak near non-interacting point.
One can also check the QAH phase is gapped and hosts non-zero Chern number in this mean-field calculation.
Another feature is that, the QAH phase seems emerge from the
boundary between stripe phase and charge density wave.
This serves as a guiding principle for finding QAH phases in our calculations
in quantum phase diagram in the main text, while we find the QAH is robust and stabilized
in the intermediate interaction regime rather than the weak interaction regime.

\end{widetext}

\end{appendices}

\end{document}